\documentclass[conference]{IEEEtran}
\IEEEoverridecommandlockouts

\usepackage{cite}
\usepackage{amsmath,amssymb,amsfonts}
\usepackage{algorithmic}
\usepackage{graphicx}
\usepackage{textcomp}
\usepackage{xcolor}
\def\BibTeX{{\rm B\kern-.05em{\sc i\kern-.025em b}\kern-.08em
    T\kern-.1667em\lower.7ex\hbox{E}\kern-.125emX}}

\usepackage{fancyvrb}
\usepackage{verbatim}
\usepackage{adjustbox}

\usepackage[english]{babel}

\usepackage{listings}

\usepackage{moresize}

\usepackage{float}

\definecolor{listinggray}{gray}{0.9}
\definecolor{lbcolor}{rgb}{1.0,1.0,1.0}
\usepackage{microtype}
\usepackage{doi}

\usepackage[subtle, charwidths=tight, leading=tight, lists=tight]{savetrees}
\usepackage{hyperref}

\usepackage{cprotect}

\begin{document}

\newcommand\idest[0]{i.\@ e.\@}

\newcommand\exempligratia[0]{e.\@ g.\@}

\newfloat{lstfloat}{htbp}{lop}
\floatname{lstfloat}{Listing}
\def\lstfloatautorefname{Listing} 

\title{Marionette: Data Structure Description and Management for Heterogeneous Computing\\
\thanks{This work is financed by Portuguese national funds through the FCT -- Fundação para a Ciência e a Tecnologia, I.P., within the scope of the projects \texttt{PRT/}\allowbreak\texttt{BD/}\allowbreak\texttt{153342/}\allowbreak\texttt{2021} (DOI: \href{https://doi.org/10.54499/PRT/BD/153342/2021}{\allowbreak\texttt{10.54499/}\allowbreak\texttt{PRT/}\allowbreak\texttt{BD/}\allowbreak\texttt{153342/}\allowbreak\texttt{2021}}), \texttt{2024.}\allowbreak\texttt{06584.}\allowbreak\texttt{CERN} (DOI: \href{https://doi.org/10.54499/2024.06584.CERN}{\texttt{10.54499/}\allowbreak\texttt{2024.}\allowbreak\texttt{06584.}\allowbreak\texttt{CERN}}) and \texttt{UIDB/}\allowbreak\texttt{50021/}\allowbreak\texttt{2020} (DOI: \href{https://doi.org/10.54499/UIDB/50021/2020}{\allowbreak\texttt{10.54499/}\allowbreak\texttt{UIDB/}\allowbreak\texttt{50021/}\allowbreak\texttt{2020}}).}
}

\author{\IEEEauthorblockN{Nuno dos Santos Fernandes}
\IEEEauthorblockA{\textit{Instituto Superior T\'{e}cnico, U. Lisboa, Portugal} \\
\textit{LIP, Lisbon, Portugal}\\
\textit{The ATLAS Collaboration, CERN, Geneva, Switzerland} \\
nuno.dos.santos.fernandes@cern.ch}
\and
\IEEEauthorblockN{Pedro Tomás}
\IEEEauthorblockA{\textit{INESC-ID} \\
\textit{Instituto Superior T\'{e}cnico}\\
\textit{Universidade de Lisboa, Portugal} \\
pedro.z.tomás@tecnico.ulisboa.pt}
\and
\IEEEauthorblockN{Nuno Roma}
\IEEEauthorblockA{\textit{INESC-ID} \\
\textit{Instituto Superior T\'{e}cnico}\\
\textit{Universidade de Lisboa, Portugal} \\
Nuno.Roma@inesc-id.pt}
\and
\IEEEauthorblockN{Frank Winklmeier}
\IEEEauthorblockA{\textit{Institute for Fundamental Science,} \\
\textit{University of Oregon, United States of America}\\
\textit{The ATLAS Collaboration, CERN, Geneva, Switzerland} \\
frank.winklmeier@cern.ch}
\and
\IEEEauthorblockN{Patricia Conde Mu\'{i}\~{n}o}
\IEEEauthorblockA{\textit{Instituto Superior T\'{e}cnico, U. Lisboa, Portugal} \\
\textit{LIP, Lisbon, Portugal}\\
\textit{The ATLAS Collaboration, CERN, Geneva, Switzerland} \\
patricia.conde.muino@cern.ch}
}

\maketitle

\begin{abstract}

Adapting large, object-oriented C++ codebases for hardware acceleration might be extremely challenging, particularly when targeting heterogeneous platforms such as GPUs. Marionette is a C++17 library designed to address this by enabling flexible, efficient, and portable data structure definitions. It decouples data layout from the description of the interface, supports multiple memory management strategies, and provides efficient data transfers and conversions across devices, all of this with minimal runtime overhead due to the compile-time nature of its abstractions. By allowing interfaces to be augmented with arbitrary functions, Marionette maintains compatibility with existing code and offers a streamlined interface that supports both straightforward and advanced use cases. This paper outlines its design,
usage, and performance, including a CUDA-based case study demonstrating its efficiency and flexibility.

\end{abstract}


\section{Introduction}

Data structure usage in a context in which heterogeneous accelerators are involved is non-trivial. Adapting a pre-existing code-base that was not written with accelerators in mind can be even more challenging, especially when object-oriented design patterns were adopted and algorithmically relevant code was written to leverage those. Since, in most cases, accelerator-centric code must coexist with some processing being done on the host, there is the additional constraint of being able to interoperate between data structures resident on either. Given that accelerators may favour different memory access patterns, the goal is supporting not only multiple ways of managing memory, but especially different ways in which to lay the data structures out in memory, while maintaining a consistent and convenient interface to the objects contained within them.

Specifically within the context of C++ and its accelerator-related extensions, the most immediate solution would be to write the desired data structures and all functionality to transfer and/or convert between them and the pre-existing host structures by hand. However, this is an onerous, bug-prone process that introduces multiple sources of truth (host data structures, accelerator data structures, host to accelerator conversions, accelerator to host conversions), which can be especially inconvenient if the original data structures themselves may be subject to changes.

Given the grammar of C++ and the (current) limitations in both reflection and programmatic code generation, there is no straightforward way to automatise the writing of all relevant classes and functions from a unified description of the intended design of the data structures. In particular, the ability to ensure that the interface of the automatically generated data structures matches the expectations of pre-existing code, such as in terms of accessors and mutators, is highly non-trivial to provide. In any case, to ensure maintainability, the solution must allow programmers of varying levels of skill and experience to engage with the mechanisms through which data structures and their interface are defined.

As a result of all of these considerations, the Marionette (short for \textbf{M}emory \textbf{A}bstracted \textbf{R}epresentation with \textbf{I}nterfaces in \textbf{O}bjects \textbf{N}ecessitating \textbf{E}xtensively \textbf{T}emplated \textbf{T}ypes \textbf{E}DM\footnote{EDM stands for ``Event Data Model'' as the project arose from an attempt to accelerate event processing at the ATLAS Experiment\cite{ATLASMain} with GPUs.}) library was developed, in an attempt to provide a convenient user experience while employing advanced compile-time programming techniques to ensure the desired behaviour can be achieved at minimal runtime cost, offering a wide variety of customisation points that more advanced users can leverage without compromising the ease of use of simpler functionality.

The code may be found in the following repository: \url{https://gitlab.cern.ch/dossantn/marionette}

\section{Context and Related Work} \label{sec:SotA}

Abstracting the description of data structures lies at a challenging intersection between requiring minimal changes to pre-existing code, supporting the broadest possible variety of environments, compilers and build systems, offering the best performance possible and minimising syntactical overhead.

Trivially, the solution of writing all the code related to data structures by hand is possible, even if cumbersome. Adding a code generation step, or transpiling to C++ from a different language or dialect, can offer the most intuitive syntax, but it may increase the complexity of the build process and raises concerns in regards to maintainability and support for present and future forms of hardware accelerators. On the other hand, relying on compiler extensions to (partially) automate the generation of the desired code, such as in \cite{Annotations}, implies a dependency on a particular subset of compilers, which may not offer support for all accelerators of interest.

Solutions that work solely at the level of C++ source code are only dependent on compiler support for the required C++ standard. Reflection and code generation functionalities are the most general way of approaching this problem. However, the current proposal to add reflection to the C++26 standard \cite{C++26Reflection} only offers limited possibilities for code generation (\Verb`define_`\allowbreak\Verb`aggregate`), and compilers and tool-chains for hardware accelerators may not always fully support the latest standards.

The basic grammar of C++ itself offers some possibilities for modifying the contents and interface of a type, through macros, templates and inheritance. Several projects take this approach, such as \cite{SoAx}, \cite{Ikra-Cpp}, \cite{TemplateBased1}, \cite{TemplateBased2} or \cite{LLAMA}, with varying support for accelerators and varying trade-offs between performance, generality and user experience. Marionette employs similar strategies, while offering increased flexibility of the defined interfaces and a wide variety of customisation points without loss of performance.

\section{Motivating Example} \label{sec:Motiv}

To guide the discussion of the design of Marionette and its functionality, one can consider a motivating example, corresponding to a simplified version of a practical use case.

\begin{lstfloat}[b]
\vspace*{-0.8cm}
\begin{minipage}[]{\linewidth}
    \begin{minipage}[]{0.49\linewidth}
\begin{lstlisting}[language={[11]C++},caption={Example \texttt{Sensor}.},label={listing:ExampleSensor}]
class Sensor {
  SensorType m_type;
  uint64_t   m_counts;
  float      m_energy;
  class Calibration {
    bool m_noisy;
    float m_parameter_A,
          m_parameter_B;
    float m_noise_A, m_noise_B;
    //Depend on the sensor type
   public:
    /* getters and setters */
  } m_calibration_data;
 public:  
  /* getters and setters */
  void calibrate_energy();
  float get_noise() const;
};
\end{lstlisting}
    \end{minipage}
    \hfill
    \begin{minipage}[]{0.49\linewidth}
\begin{lstlisting}[language={[11]C++},numbers=none,caption={Example \texttt{Particle}.},label={listing:ExampleParticle}]
class Particle {
  float    m_energy;
  float    m_x, m_y;
  uint64_t m_origin;
  std::vector<uint64_t>
           m_sensors; 
  float    m_x_variance,
           m_y_variance;
  float    m_significance
          [SensorType::Num];
  float    m_E_contribution
          [SensorType::Num];
  uint8_t  m_noisy_count
          [SensorType::Num];
  //Separate for each sensor type
 public:
  /* getters and setters */
};
\end{lstlisting}
    \end{minipage}
    \hfill
\end{minipage}
\vspace*{-0.5cm}
\end{lstfloat}
Drawing from the context of its original design, suppose there is a 2D grid of sensors (of different types) that measure the energy of incoming particles, with a set of calibrated parameters expressing the conversion from raw sensor counts to energy and an estimate of the noise associated with the measurement. The particles are reconstructed based on the energy measurements, including a list of sensors that contributed to the reconstruction and properties that are tracked separately for each type of sensor. Both sensor and particle information will be expressed in a mostly object-oriented way, as shown in listings \ref{listing:ExampleSensor} and \ref{listing:ExampleParticle}.

Suppose, then, that part of the code interacting with these structures will be adapted to use hardware acceleration, but, due to more complicated dependencies and/or personpower limitations, a non-negligible portion of the algorithms will have to be ported gradually, if at all. Thus, code referencing the \Verb`Sensor` and \Verb`Particle` class must remain valid on both host and accelerator. In addition, the ability to experiment with different data layouts may be useful for development efforts and optimization.

Most of the discussion will follow from this example, including the benchmarks presented in section \ref{sec:CaStu}.

\section{Main Requirements}
As suggested in the previous sections, Marionette was developed to offer a general solution for abstracting the layout of data structures in an accelerator-compatible way. The range of supported heterogeneous computing platforms should be as broad as possible. Given current trends, GPUs would be the most natural use case, but the design should not exclude support for FPGAs and other forms of co-processor. The set of requirements for the design of Marionette may be summarized thus:

\begin{itemize}
	\item Allow the description of data structures with unified interface across multiple heterogeneous computing platforms:
		\begin{itemize}
			\item Handle data structure generation at compile time, avoiding unnecessary run-time overhead and polymorphism;
			\item Provide an intuitive object-oriented interface on top of the underlying data structures (array-of-structures-like);
			\item Support user-provided data storage strategies;
			\item Provide basic data storage strategies that can be readily adapted to new heterogeneous computing devices by specifying a minimal set of memory-related operations;
		\end{itemize}
	
	\item Implement efficient data transfers between data structures that are laid out according to different storage strategies:
		\begin{itemize}
			\item Allow the end user to specify additional data transfers, including from types outside the library;
		\end{itemize}

	\item Allow the addition of arbitrary functions to the final interface of the data structures;

	\item Support several kinds of properties within the data structures, with different semantics, which should be arbitrarily nestable;

	\item Impose no further requirements than the ability to compile C++17 for the target platform/device, but support later features when available (\exempligratia{} the spaceship operator \Verb`<=>`);
	
	\item Achieve all of this with minimal run-time impact compared to the equivalent handwritten solution.
\end{itemize}

\section{The Basic Design of Marionette}

The Marionette library is designed around two main classes: \Verb`Object` and \Verb`Collection`. These represent, respectively, a single entity with the specified contents and interface and a list (with an \Verb`std::vector`-like interface) of such structures. The three template parameters of these classes provide the core of the flexibility offered by the library.

The first template parameter corresponds to the \textit{layout}, which determines the way in which the data will be stored in memory. For objects, the layout is used internally to support both owning objects and proxies into collections (which provide the object-oriented interface of the data structures). For collections, it is specified by the user according to their needs. For example, allocating memory in blocks of a given size, as opposed to a pure structure-of-arrays where each array is contiguous, or on a hardware accelerator as opposed to the host. Sub-section \ref{ssec:DefLay} provides more details on the layouts offered by Marionette and on how to define a new layout.

The second template parameter corresponds to a compile-time list of \textit{properties} to be held by the collection. This allows generating the data structures according to the layout, with the desired interface. A property, in this sense, is a generalisation of a member variable to be associated with each object. Defining properties will correspond to the majority of end-user interactions with the Marionette framework, as explained in the following section.

The third template parameter holds \textit{meta-information} that provides a more convenient interface to certain kinds of properties. As this is managed internally by the library, the end user should simply rely on the default value of the template parameter when declaring variables (\idest{} not provide it explicitly).

\section{Implementing Data Structures in Marionette}
Properties are represented by a \textit{property description class}, which expresses compile-time information about the desired semantics of the property and any other relevant information (\exempligratia{} the type of a variable). The property description class also specifies functions that will be added (via inheritance) to a collection or object that contains the property in question. These would be implemented in \Verb`CollectionFunctions` and \Verb`ObjectFunctions` member types, which are templated on the type of the overall collection or object and the underlying layout. Within those functions, casting the \Verb`this` pointer to a pointer to the final class allows accessing the full interface of the \Verb`Collection` or \Verb`Object`, including other functions associated with the properties. Of note are \Verb`get_collection`, \Verb`get_object` and \Verb`get`, which access the storage corresponding to a property description class.

The most common kind of property would be a single native type that is associated with every object in a collection: a \textbf{\textit{per-item property}}. For this, one must specify the type to be stored, as well as any functions (\exempligratia{} accessors and mutators) that provide the desired interface. To give a concrete example, listing \ref{list:explicit} implements the \Verb`energy` property of a \Verb`Sensor`.

Given that this is a very common pattern, a more convenient \Verb`MARIONETTE_`\allowbreak\Verb`DECLARE_`\allowbreak\Verb`PER_`\allowbreak\Verb`ITEM_`\allowbreak\Verb`PROPERTY` macro is provided, taking as arguments the uncapitalized and capitalized forms of the property name, with the latter being used as the name of the property description class, and the type to be stored. Several ways are offered to customise the generation of the accessor and mutator names. Similar property declaration macros exist for most other kinds of properties, all of the form \Verb`MARIONETTE_`\allowbreak\Verb`DECLARE_*` and taking the names as the first two arguments.

\begin{lstfloat}
\vspace*{-0.5cm}
\begin{lstlisting}[language={[11]C++},caption={Implementing the \texttt{energy} of a \texttt{Sensor} explicitly.},label={list:explicit}]
struct Energy : Marionette::InterfaceDescription::PerItemProperty {
 using Type = float;
 template <class F, class L> struct ObjectFunctions {
  const float & energy() const {
   return static_cast<const F*>(this)->template get<Energy>(); }
  float & energy() {
   return static_cast<F*>(this)->template get<Energy>(); }
  void set_energy(const float & e) {
   static_cast<F*>(this)->template get<Energy>() = e; }
 };
 template <class F, class L> struct CollectionFunctions {
  decltype(auto) energy() const {
   return static_cast<const F*>(this)->template get_collection<Energy>(); }
  decltype(auto) energy() {
   return static_cast<F*>(this)->template get_collection<Energy>(); }
  template <class T> void set_energy(T && t) {
   static_cast<F*>(this)->energy() = std::forward<T>(t); }
  decltype(auto) energy(const typename Layout::size_type i) const {
   return static_cast<const F*>(this)->template get_object<Energy>(i); }
  decltype(auto) energy(const typename Layout::size_type i) {
   return static_cast<F*>(this)->template get_object<Energy>(i); }
  void set_energy(const typename Layout::size_type i, const float & e) {
   static_cast<F*>(this)->energy(i) = e; }
}; };
\end{lstlisting}
\vspace*{-0.5cm}
\end{lstfloat}

\begin{lstfloat}
\vspace*{-0.4cm}
\begin{lstlisting}[language={[11]C++},caption={Implementing \texttt{Sensor} and \texttt{Particle} in Marionette.},label={listing:Marionettification}]
using Marionette::InterfaceDescription::PropertyList;
MARIONETTE_DECLARE_PER_ITEM_PROPERTY(type, Type, SensorType);
MARIONETTE_DECLARE_PER_ITEM_PROPERTY(counts, Counts, uint64_t);
MARIONETTE_DECLARE_PER_ITEM_PROPERTY(energy, Energy, float);
MARIONETTE_DECLARE_PER_ITEM_PROPERTY(noisy, Noisy, bool);
MARIONETTE_DECLARE_PER_ITEM_PROPERTY(parameter_A, ParameterA, float);
MARIONETTE_DECLARE_PER_ITEM_PROPERTY(parameter_B, ParameterB, float);
MARIONETTE_DECLARE_PER_ITEM_PROPERTY(noise_A, NoiseA, float);
MARIONETTE_DECLARE_PER_ITEM_PROPERTY(noise_B, NoiseB, float);
MARIONETTE_DECLARE_SUBGROUP_PROPERTY(calibration_data, CalibrationData,
                            Noisy, ParameterA, ParameterB, NoiseA, NoiseB);
struct SensorFuncs : Marionette::InterfaceDescription::NoProperty;
//Defines appropriate calibrate_energy and get_noise object functions
using Sensor = PropertyList<Type, Counts, Energy, CalibrationData, SensorFuncs>;
MARIONETTE_DECLARE_PER_ITEM_PROPERTY(x, X, float);
MARIONETTE_DECLARE_PER_ITEM_PROPERTY(y, Y, float);
MARIONETTE_DECLARE_PER_ITEM_PROPERTY(origin, Origin, uint64_t);
MARIONETTE_DECLARE_SIMPLE_JAGGED_VECTOR(sensors, Sensors, int, uint64_t);
MARIONETTE_DECLARE_PER_ITEM_PROPERTY(x_variance, XVariance, float);
MARIONETTE_DECLARE_PER_ITEM_PROPERTY(y_variance, YVariance, float);
MARIONETTE_DECLARE_SIMPLE_ARRAY_PROPERTY(significance, Significance,
                                         SensorType::Num, float);
MARIONETTE_DECLARE_SIMPLE_ARRAY_PROPERTY(E_contribution, EContribution,
                                         SensorType::Num, float);
MARIONETTE_DECLARE_SIMPLE_ARRAY_PROPERTY(noisy_count, NoisyCount,
                                         SensorType::Num, uint8_t);
using Particle = PropertyList<Energy, X, Y, Origin, Sensors,
                              XVariance, YVariance, Significance,
                              EContribution, NoisyCount>;
\end{lstlisting}
\vspace*{-0.9cm}
\end{lstfloat}

A \textbf{\textit{no-property} property} allows extending the interface of collections or objects without any associated storage. The property description class must inherit from \Verb`NoProperty` and define the \Verb`ObjectFunctions` and \Verb`CollectionFunctions` classes. This would be used to implement the \Verb`calibrate_`\allowbreak\Verb`energy` and  \Verb`get_`\allowbreak\Verb`noise` member functions of a \Verb`Sensor`.

To reproduce the semantics given in the example for \Verb`calibration_data`, one would employ a \textbf{\textit{sub-group property}}. The property description class inherits from \Verb`SubGroup` and defines a \Verb`Properties` member type corresponding to a compile-time list of other property descriptions to be contained within the sub-object. These will be treated as separate properties in the same way as if they were declared at the top level (\exempligratia{} being stored as separate arrays in a structure-of-arrays layout). The property declaration macro will take the list of other properties starting from its third argument.

The members of the example \Verb`Particle` that are tracked by sensor type could benefit from being stored in separate arrays for each type, while still providing the interface of an array within each object. This would be an \textbf{\textit{array property}}, which in essence has both a ``vector of arrays'' and an ``array of vectors'' interface.  The property description class will contain the \Verb`Properties` to be held in the array and the compile-time \textit{extent} of the array. Besides the expected property declaration macro, which takes the extent and then the list of properties starting from its fourth argument, a \Verb`*_SIMPLE_*` version will take just the extent and a type to cater to the common case where a single per-item property is held. 

A \textbf{\textit{jagged vector property}} allows associating a dynamic number of values to each object. The set of all the values for all the objects in a collection is accessible as if it were a single, continuous vector. To represent this, the prefix sum of the sizes of each individual vector is stored in a \textit{global property}, which is not added to the interface of individual objects. In addition, the concept of a \textit{size tag} must be introduced so that differently sized arrays may coexist within a collection, as the set of all values for all the objects has a size that is independent from the overall number of objects. This handled in a way that is transparent to the end user. The property description class specifies the type to be used to represent the prefix sum of the sizes of the jagged vectors (which may be smaller than the \Verb`size_type` of the collection) and the list of properties to be held. Similarly to array properties, a \Verb`*_SIMPLE_*` version of the macro handles the case where there is a single per-item property within the vectors, as in the \Verb`sensors` of the \Verb`Particle`.

A possible implementation of the \Verb`Sensor` and \Verb`Particle` classes within Marionette is shown in listing \ref{listing:Marionettification}.

\section{Advanced Usage}

\subsection{Support for Heterogeneous Computing} \label{ssec:SuHeCo}

Marionette employs two main abstractions to support hardware accelerators: \textit{execution contexts}, which abstract where the code is being compiled for (\exempligratia{} ``CPU'', ``GPU with CUDA'', ``GPU with HIP'' or ``GPU with OpenCL''), and \textit{memory contexts}, which encapsulate a way of managing memory. While support for new execution contexts may require some changes to the internals, users are free to define new memory contexts.

Each memory context has an associated \Verb`ContextInfo` type to hold runtime information related to each individual allocation.
Memory contexts specify how to \Verb`allocate`, \Verb`deallocate` and \Verb`memset` memory, according to the provided context information. By design, every collection will have an associated memory context, determined by the collection's layout, and store the corresponding context information. An \Verb`update_memory_context_info` member function of \Verb`Collection` is offered to allow modifications of the context information after a collection has been constructed. This can even mean allocating memory using the new information, copying from the old memory and deallocating it.

Transferring memory between different contexts is made possible through \Verb`memcopy_with_context` and several variants of it that support partially overlapping arrays for easier insertion and deletion. These are defined as free functions, in order to support several combinations of contexts. Depending on the memory contexts in question, additional parameters may be provided to control some aspects of the data transfer.

In order to ensure all the relevant functions will be callable from all execution contexts, a set of \Verb`MARIONETTE_`\allowbreak\Verb`PORTABILITY_`\allowbreak\Verb`ANNOTATIONS` is defined and used to decorate class and function declarations throughout the internals of Marionette. These should expand to the appropriate annotations (\exempligratia{} \Verb`__host__ __device__`), with the user being able to extend or override their definition. User-defined functions and classes may employ them as well, for maximum portability.

\subsection{Defining Layouts} \label{ssec:DefLay}

The most powerful feature of Marionette is the ability to define different ways of storing the data by writing a \textit{layout class}. The actual storage is described in the \Verb`layout_holder` member type, which is templated on three parameters: the compile-time list of properties to be stored and two multiplicative factors to the extent of the properties and size tags, which are needed to fully support properties being nested within array properties.

The layout class must specify the \Verb`size_type` and \Verb`difference_type` to be used for indexing within the collection, the memory context associated with the collection and \Verb`interface_properties` to specify what can be done with its contents (which may vary with the execution context). Other optional member functions may also be defined in order to enable optimizations and cater to specific use cases.

Implementing the \Verb`layout_holder` requires a certain amount of technical complexity since any combination of user properties must be properly supported. The required interface corresponds to a set of simple size-changing operations (\Verb`resize`, \Verb`reserve`, \Verb`clear`, \Verb`shrink_to_fit`, \Verb`insert` and \Verb`erase`), which will act on either individual ``arrays'' or the entirety of the properties under a given size tag, as well as access to these same ``arrays'' and size tags. The ``arrays'' need not be contiguous, only a mapping from an index to a variable of the required type.

Marionette provides out-of-the-box support for use cases where a vector-like container should be associated with each per-item property (the \Verb`VectorLikePerProperty` layout) and where a single block of memory, corresponding to a maximum size that may be specified for each property, is allocated using an allocator-like class (the \Verb`DynamicStruct` layout). Some template parameters provide additional flexibility. Accelerators are immediately supported by using appropriate containers or allocators, such as the provided \Verb`ContextAwareAllocator` and the \Verb`ContextAwareVector` classes, which rely on the functions defined in the memory context. This means that supporting new accelerators simply requires having an appropriate memory context to represent allocations on the accelerator.

Transferring data between layouts is exposed both through normal copy/move assignment and \Verb`copy` and \Verb`move` functions, which may receive the same additional arguments as a \Verb`memcopy_with_context` between the contexts of the two collections. A comprehensive set of defaults that copy the arrays corresponding to each property one by one are provided, but users may provide more optimal implementations or specify transfers from pre-existing data structures defined outside of Marionette through appropriate specialisations of the \Verb`TransferSpecification` class. This is templated on a \Verb`TransferPriority` enumerate that allows gracefully falling back to more general implementations.

\section{Case Studies}\label{sec:CaStu}

The first prototypes obtained by applying Marionette within its original use case of accelerating the processing of events at the ATLAS Experiment with GPUs show that the generated PTX code matches the handwritten solution. The same happens for some of the tests included in the Marionette repository, such as \Verb`GPU_{nth}_test.cu`. In other words, the abstractions offered by Marionette have no runtime impact in these cases.

A more detailed study can nonetheless be carried out based on the example described in section \ref{sec:Motiv}, whose source may be found in the \Verb`realistic_example.cu` test in the repository. This represents a realistic scenario where the data structures may be used:  raw counts are used to calculate the energy of each sensor, particles are detected based on the $5\times5$ neighbourhood around a sufficiently energetic sensor, and the remaining properties are calculated from the sensors that contributed in that neighbourhood. The benchmarks focus on demonstrating that the usage of Marionette has no impact compared to the equivalent hand-written solution for using a structure-of-arrays, both on the host and using CUDA for GPU acceleration. They were obtained on a system with an Intel Xeon Gold 5220 and an NVidia RTX A6000. The time measurements come from the average of the ten fastest times out of 50 executions of 10 different events.

\begin{figure}[h]
\vspace*{-0.3cm}
\centerline{\includegraphics[width=0.673\columnwidth]{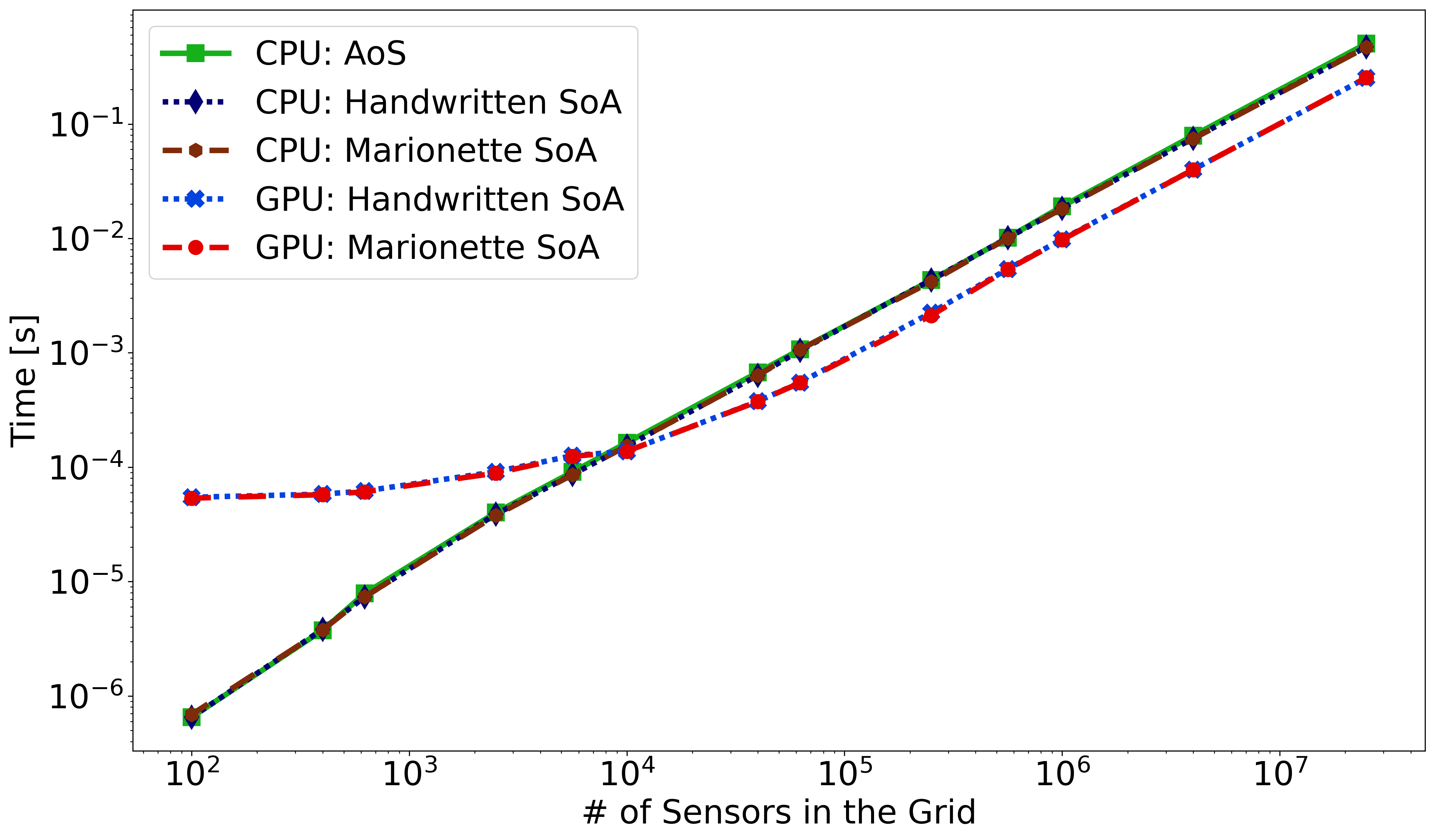}}
\vspace*{-0.2cm}
\caption{Execution time for sensor-related computations in the realistic test.}
\label{fig:sensortime}
\vspace*{-0.3cm}
\end{figure}

Figure \ref{fig:sensortime} shows the time taken to fill the data structures with the raw sensor information, transfer it to the GPU (if applicable) and calculate the sensor energy, as a function of the number of sensors in the grid. The overheads associated with GPU acceleration outweigh any gains for a grid smaller than $100\times100$, after which the difference between CPU and GPU execution times is fixed, suggesting that the data transfers and conversions are dominant. There also seems to be no difference between array-of-structures and structure-of-arrays on the CPU, likely due to the fact that the majority of the information in the \texttt{Sensor} class is needed for the computation. In any case, for both CPU and GPU operation, Marionette and the equivalent handwritten solution display exactly the same performance.

\begin{figure}[h]
\vspace*{-0.3cm}
\centerline{\includegraphics[width=0.673\columnwidth]{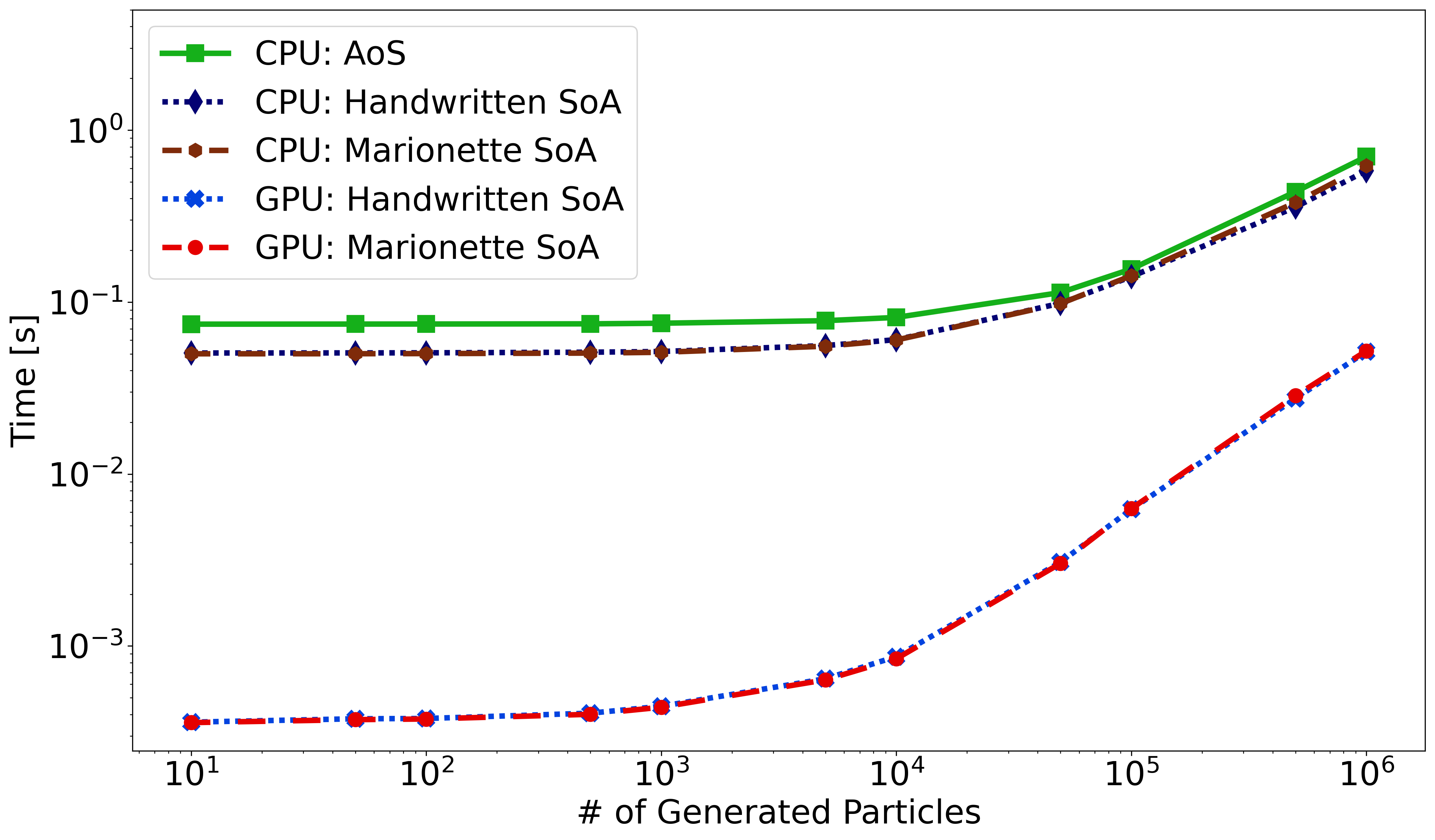}}
\vspace*{-0.2cm}
\caption{Execution time for particle-related computations in the realistic test.}
\label{fig:particletime}
\vspace*{-0.15cm}
\end{figure}

On the other hand, figure \ref{fig:particletime} shows the time taken to reconstruct the particles, transfer back to the CPU (if applicable) and fill back the original array-of-structures, as a function of the number of particles that were generated, for a grid of $5000\times5000$ sensors. In this case, there is a clear speed-up from GPU acceleration, though the overheads of data transfers and conversions become increasingly relevant as the number of particles goes above $10^4$. This is particularly evident in the CPU case, for which the performance advantage of a structure-of-arrays layout becomes less significant. Once again, there are no differences in performance between Marionette and the equivalent handwritten solution, both on the CPU or the GPU, confirming the zero-cost nature of the library's abstractions.

Data structures with a much more complex interface can be implemented within Marionette. Given the strictly compile-time nature of the mechanisms employed to generate the data structures, the expectation is that, in every case, the resulting code will be optimised to be equivalent to the (much more cumbersome) handwritten solution. In general, the additional possibility of sidestepping unnecessary conversions (in this case, the final step of filling the pre-existing data structures) can bring even more benefits, together with the possibility of fine-tuning the layout to the needs of the problem without loss of generality or ergonomy of the final code.

\section{Summary and Conclusions}

The Marionette library offers a comprehensive set of abstractions over the description of data structures that allows supporting multiple memory allocation strategies while maintaining the same interface. The ability to specify arbitrary functions as part of the interface of the data structures or each individual object contained within them allows replicating the behaviour of pre-existing code, without incurring any run-time costs since everything is resolved during compilation. The many possibilities for customisation of the behaviour allow advanced users to fulfil the particular requirements of their use case, without compromising ease of use for more common cases.

Applying Marionette to some concrete use cases demonstrates the zero-cost nature of the abstractions it provides, as code using Marionette data structures will compile to the same operations as the equivalent hand-written data structures. Work is ongoing to employ Marionette within the more complex use case that motivated its design, with preliminary results suggesting that the zero-cost nature of the abstractions will remain.

Several other improvements to the functionality of the library are foreseen, including new types of properties, new layouts and support for more heterogeneous computing platforms.

\bibliographystyle{IEEEtran}
\bibliography{IEEEabrv, bibliography.bib}

\end{document}